\documentclass[12pt]{article}
\usepackage{caption}
\usepackage{graphicx} 
\usepackage{amssymb}
   \title{$\nu$ production in Centaurus A and M87 from $\gamma$-ray 
   interactions with the gas and dust at the sources}
   \author{J.C. Arteaga-Vel\'azquez}
   \date{}

\begin{document}
 
   \maketitle
   \vspace{-3pc}
   \begin{center}
   \textit{Instituto de F\'\i sica y Matem\'aticas, Universidad Michoacana,
    Edificio C3, Cd. Universitaria, 58040 Morelia, Michoacan, Mexico}\\
    Email: arteaga@ifm.umich.mx
   \end{center}

   \begin{abstract}     
     Centaurus A and M87 are the closest galaxies with active galactic nuclei and
     TeV $\gamma$-ray emission. The existence of such TeV radiation suggests the
     production of a neutrino flux from the photo-hadronic interactions of the 
     $\gamma$-photons of the active galaxies and their own gas and dust content. 
     Using a simple model of Centaurus A and M87, the corresponding $\nu$ luminosities 
     at source and their fluxes at Earth were calculated. The neutrino fluxes associated 
     with the aforementioned process resulted to be  $E^2 \Phi_{\nu + \bar{\nu}} \, 
     \lesssim 10^{-13} \,\,  \mbox{s}^{-1} \, \, \mbox{GeV} \, \mbox{cm}^{-2}$, more than 
     $6$ orders of magnitude below the modern upper limits from neutrino  
     telescopes. It will be shown that, at high-energies relevant for neutrino 
     astronomy, these $\nu$ fluxes are not competitive with those fluxes which could 
     be produced from astrophysical scenarios involving cosmic ray acceleration.
   \end{abstract}

   \vspace{1pc}

   \vspace{1pc}
   \textit{This is an author-created, un-copyedited version of an article published in Journal 
   of Physics: Conference Series. IOP Publishing Ltd is not responsible for any errors 
   or omissions in this version of the manuscript or any version derived from it. The 
   definitive published authenticated version is available online at 
   doi:10.1088/1742-6596/378/1/012005.}

   \section{Introduction}

   Centaurus A and M87 are two giant elliptical radio galaxies, which are classified as 
   Fanaroff-Riley I (FR I) type sources \cite{FRI}. Centaurus A, also known as NGC 5128, 
   is located at a distance of $3.4 \, \mbox{Mpc}$ from the Earth  \cite{Quillen06}. 
   Meanwhile, M87 is some $16.7 \, \mbox{Mpc}$ away \cite{m87-multi09}. These galaxies 
   host the closest active galactic nuclei to the Earth and are the sources of high-energy 
   radiation in the form of TeV $\gamma$-rays \cite{centA-egret98}-\cite{m87-fermi09}. 
   Recently, the Pierre Auger observatory (PAO) detected some ultrahigh-energy 
   cosmic ray (UHECR) events within $3.1^\circ$ around the direction of Centaurus A 
   \cite{Auger07}. Besides, the PAO showed a weak correlation between
   the arrival direction of cosmic rays with energies above $6 \times 10^{19} \, 
   \mbox{eV}$ and the distribution of AGN's \cite{Auger07}. Both PAO's results 
   suggest, in this way, that Centaurus A and M87 could also be the production
   place of UHECR.

   The presence of high-energy gamma- and cosmic rays in Centaurus A and M87 
   would imply the existence of an associated flux of high-energy neutrinos 
   produced from the interactions of both types of high-energy radiation with 
   the ambient gas and photons of the corresponding galaxies. For the case
   of UHECR, $\nu$ production mediated by pion photoproduction has been widely 
   studied in the literature (see, for example, \cite{Kachelriess09, Hylke08, 
   halzen08, halzen11}) and its contribution is though to be the main 
   mechanism of $\nu$ emission. However, it remains to be proved beyond any 
   reasonable doubt that UHECR are produced in Centaurus A and M87 to guarantee 
   the production of neutrinos from the above process. On the other hand, 
   although it is known that several AGN's are $\gamma$-ray emitters, the 
   mechanisms of neutrino production induced by $\gamma$-ray collisions 
   \cite{Razzaque06, Menon09, Beaudet66} in that context are scarcely studied, 
   mainly because they are not expected to produce important $\nu$ fluxes 
   due to their low cross sections. In any case, these secondary processes should 
   be also contributing to the production of high-energy neutrinos in active 
   galaxies. In the present work, one of such $\nu$ production mechanisms is 
   studied: the photo-hadronic interaction of gamma-rays with the gas and dust 
   content of active galaxies, in particular, for NGC5128 and M87, assuming 
   that the gamma sources are located at the core of their AGN's \footnote{The 
   exact position of the high-energy source inside active 
   galaxies is still an open problem. One candidate is the core of AGN's. In 
   fact, multi-wavelength observations of M87 point out that the TeV gamma-ray 
   production site could be located at the nucleus of its AGN \cite{m87-multi09}.}.
   Accordingly, the corresponding $\nu$ fluxes are estimated and compared with 
   those calculated in the framework of some astrophysical models involving 
   UHECR.

   \section{The neutrino induced flux}

   Let's be $L_\gamma^{*} (\epsilon_\gamma, r) = dN_\gamma/dt\,d\epsilon_\gamma$ 
   the photon spectral luminosity of the AGN measured  in the source frame 
   for a photon energy $\epsilon_\gamma$ at a distance $r$ from the core. 
   Let's also assume that the gas and dust of the host galaxy are found in the 
   form of protons and that their kinetic energies are negligible in
   comparison with the energies of the gamma-rays (supposition that can work 
   for the hot, warm and cold gasses of the interstellar medium
   \footnote{For example, the temperature of the gas supply for the central 
   engine of some AGN's has been estimated to be of the order of $\mbox{keV}$ 
   \cite{Allen06, Evans06, centA-Evans04}, in such a medium, the mean kinetic
   energy expected for protons is roughly of the same order of magnitude, 
   which results to be small in comparison with the gamma energies considered
   in this work ($\epsilon_{\gamma} > 10^{-0.8} \, \mbox{GeV}$).}). 

   In this way, the neutrino spectral luminosity, $L_\nu(\epsilon_\nu)$
   induced by the interaction of the gamma-photons with the intergalactic
   material of the galaxy on their  way out along the direction of the
   observer is given by   
   \begin{equation}
     L_\nu(\epsilon_\nu) d\epsilon_\nu = 
     \int_0^{r_f} \int_{\epsilon_{\gamma, i}}^{\epsilon_{\gamma, f}} \rho_{\small H}(r)
     \sigma_{\gamma P}(\epsilon_\gamma) 
     Y^{\gamma P \rightarrow \nu}(\epsilon_\gamma, \epsilon_\nu)
     L_\gamma^{*} (\epsilon_\gamma, r) d\epsilon_\gamma dr,
     \label{eq1}
   \end{equation} 
   where $\rho_{\small H}$ is the density of target protons at the distance
   $r$,  $\sigma_{\gamma P}$  is the $\gamma P$ cross section at a photon energy 
   $\epsilon_\gamma$  and $Y^{\gamma P \rightarrow \nu}(\epsilon_\gamma, 
   \epsilon_\nu)$ is the neutrino yield, which is defined as the number of 
   $\nu$'s with energy in the interval $d\epsilon_\nu$ around $\epsilon_\nu$  
   produced by a photon with energy in the range $[\epsilon_\gamma, 
   \epsilon_\gamma + d\epsilon_\gamma]$ after a collision with a
   proton at rest. Energy losses of the parent particles in the 
   $\nu$-production chain have been neglected.

    Since, $L_\gamma^{*}(\epsilon_\gamma, r)$ is unknown at the  core of the
    AGN, the gamma-ray luminosity at source derived from Earth  observations, 
    $L_\gamma(\epsilon_\gamma)$, will be used instead in equation
    (\ref{eq1}). Note that by doing so the estimated $\nu$ flux could be
    lower than the real one, since the actual $\gamma$-ray luminosity could be 
    bigger than $L_\gamma(\epsilon_\gamma)$ at the interior of the AGN
    \cite{Ostap}. Now, by using the above approximation, equation (\ref{eq1})
    is reduced to the following expression:
   \begin{equation}
     L_\nu(\epsilon_\nu) d\epsilon_\nu = 
     \Sigma_{\small H} \int_{\epsilon_{\gamma, i}}^{\epsilon_{\gamma, f}}
     \sigma_{\gamma P}(\epsilon_\gamma) 
     Y^{\gamma P \rightarrow \nu}(\epsilon_\gamma, \epsilon_\nu)
     L_\gamma (\epsilon_\gamma) d\epsilon_\gamma,
     \label{eq2}
   \end{equation} 
   with
   \begin{equation}
     \Sigma_{\small H}  =  \int_0^{r_f}   \rho_{\small H}(r)  dr.
   \label{eq3}    
   \end{equation}  
   Here, $\Sigma_{\small H}$ is the column density of target protons along
   the line of sight from $r = 0$ up to $r_f$, which corresponds to the 
   size of the halo of the active galaxy. Integration limits for the 
   energy parameters are defined as follows: 
   $(\epsilon_{\gamma, i}, \epsilon_{\gamma, f}) = 
   (10^{-0.8}  \, \mbox{GeV}, 10^{6} \, \mbox{GeV})$ 
   and 
   $(\epsilon_{\nu, i}, \epsilon_{\nu, f}) = 
    10^{-5}  \, \mbox{GeV}, 10^{6} \, \mbox{GeV})$. 
   The lower limit on the gamma-energies is set just at the 
   threshold for pion photoproduction, $\epsilon_{\gamma, th} = 
    m_\pi(m_\pi c^2 + 2m_p c^2)/2m_p \approx 10^{-0.8} \, \mbox{GeV}$ 
   \cite{Propagation-uhecr}. Meanwhile, the upper limit is fixed 
   at $10^{6} \, \mbox{GeV}$, considering that the gamma-ray emission 
   could be of leptonic origin and therefore that it could have a cut 
   at the high-energy regime, somewhere around $10^{2} \, \mbox{TeV}$ 
   (see, for example, \cite{Reynoso10}).

   The flux of neutrinos at Earth is estimated from the $\nu$ 
   luminosity of the active galaxy by means of the formula 
  \begin{equation}
    \Phi_{\nu + \bar{\nu}}(\epsilon'_\nu) = \frac{(1+z)^2}{4\pi D^2_L} \cdot
    L_{\nu + \bar{\nu}}(\epsilon_\nu)
   \label{eq4}
  \end{equation}
  where $z$ is the redshift of the galaxy and $\epsilon'_\nu = 
  \epsilon_\nu/(1 + z)$ is the neutrino energy after redshift energy
  losses\footnote{Primed quantities are measured at Earth.}. $D_L$ is the
  luminosity distance to the source, which is defined as
  \begin{equation}
    D_L(z)  = (1 + z) \int_{0}^{z} c\,dz/H(z).
    \label{eq5}
  \end{equation} 
  Here, $c$ is the speed of light in vacuum and $H(z)$ is the Hubble 
  parameter at redshift $z$. For a cosmological scenario
  dominated by vacuum energy, with $\Omega_\Lambda = 1 - \Omega_m$,
  \begin{equation}
      c/H(z) = (c/H_0) [1 - \Omega_m + \Omega_m (1 + z)^3]^{-1/2}.
  \label{eq6}
  \end{equation} 
  Along this paper, the following parameters will be used: $\Omega_\Lambda =  
  0.74$ and $c/H_0 = 1.28 \times 10^{28} \, \mbox{cm}$  \cite{PDG10}.

  For the final fluxes of individual $\nu$ flavors, neutrino oscillations 
  will be taken into account \cite{nakamura2010}. That implies that after a 
  long way to the Earth, neutrinos of different types will be present in 
  the total flux in a $(\nu_e + \bar{\nu}_e)$:$(\nu_\mu + \bar{\nu}_\mu)$:
  $(\nu_\tau + \bar{\nu}_\tau)= 1:1:1$ ratio.

  \section{Gamma ray luminosities}

  Several gamma-ray observatories have measured the differential photon 
  spectra of Centaurus A and M87 in the MeV-TeV regime. For Centaurus A, 
  measurements of the $\gamma$-ray flux in the interval $0.1 - 1 \, \mbox{GeV}$ 
  have been performed with the EGRET detector on board of the Compton 
  Gamma-Ray Observatory \cite{centA-egret98, centA-egret99}, while in the 
  $0.1 - 30  \, \mbox{GeV}$ energy range, data has been collected 
  with the \textit{Fermi}-LAT instrument \cite{centA-fermi10}. Besides in
  the very high-energy region ($\epsilon'_\gamma >100 \, \mbox{GeV}$), 
  data has been registered with the telescopes of the H.E.S.S. experiment
  \cite{centA-hess09}. In case of M87, its gamma-ray flux has been measured 
  with the \textit{Fermi}-LAT telescope in the range from $0.2$ to $32 \, 
  \mbox{GeV}$ and with HEGRA \cite{m87-hegra}, HESS \cite{m87-hess} and  
  VERITAS \cite{m87-veritas08} in the energy regime of $0.1 - 30 \, \mbox{TeV}$.

  Using the above observational data, the $\gamma$ luminosities of the sources 
  can be easily estimated. Let's be $\Phi_\gamma(\epsilon'_\gamma) = dN_\gamma/dt'
  dA' d\epsilon'_\gamma $ the differential photon spectrum (in units of
  photons per unit time, unit area and interval of energy)  detected at Earth
  from the active galaxy. Assuming a steady and isotropic emission, the photon
  spectral luminosity of the AGN is calculated by using
  \begin{equation}
   L_\gamma (\epsilon_\gamma) = 4\pi D^2_L \cdot
   \frac{\Phi_\gamma(\epsilon_\gamma)}{(1+z)^2} \cdot 
   \mbox{e}^{\tau(\epsilon_\gamma, z)},
   \label{eq7}
  \end{equation} 
  where corrections for both adiabatic energy losses due to the redshift and 
  attenuation due to interactions with the background radiation are 
  considered. The latter is done introducing the exponential term, where
  the parameter $\tau(\epsilon_\gamma, z)$ appears. Here, $\tau$ represents
  the $\gamma \gamma$ optical depth for a photon traveling from the source 
  with initial energy $\epsilon_\gamma$. The values used for $\tau$
  were taken from reference \cite{gg-optical-depth}.

\begin{figure}[!t]
    \centering
    \includegraphics[width=77mm, height=56mm]{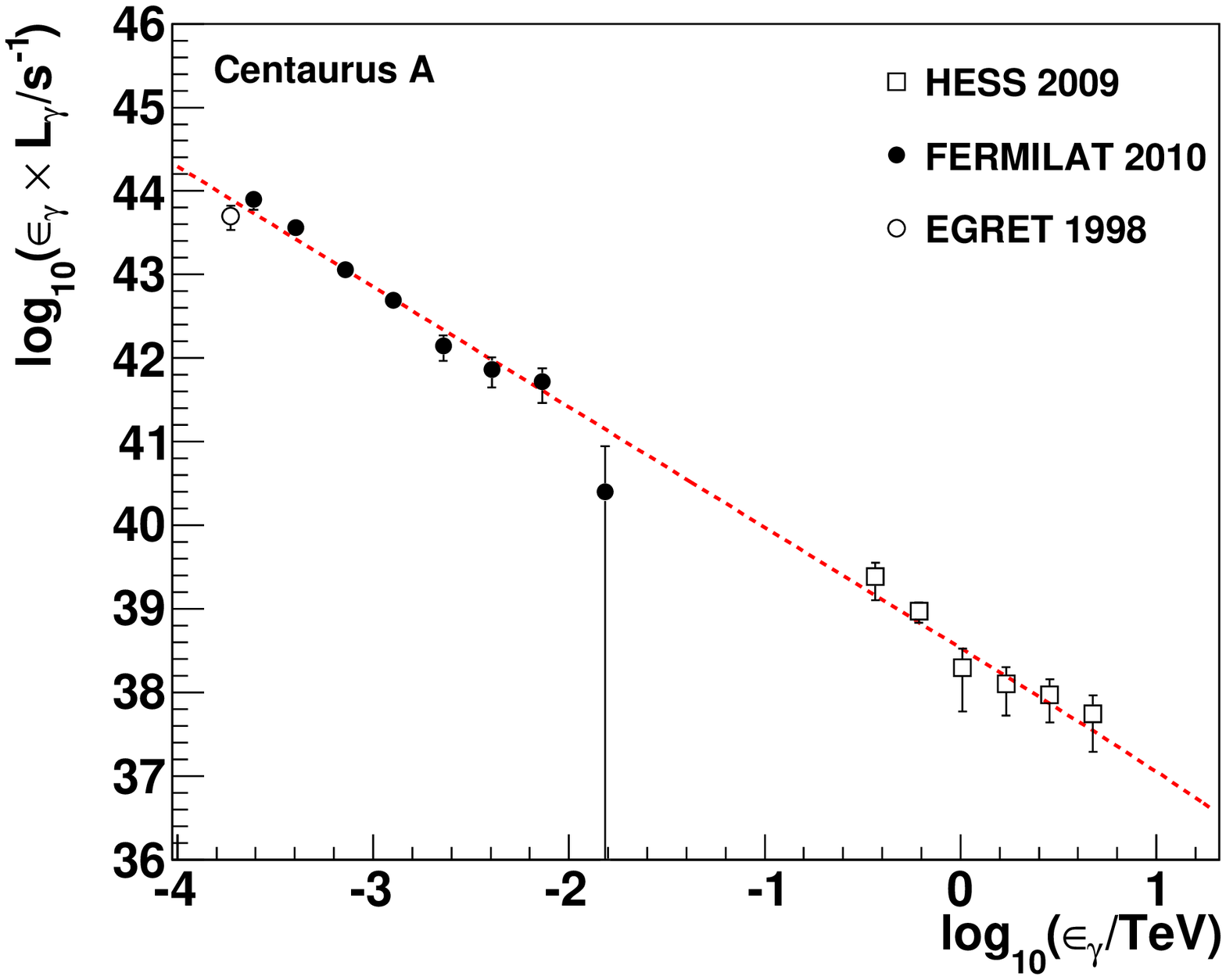}
    \includegraphics[width=77mm, height=56mm]{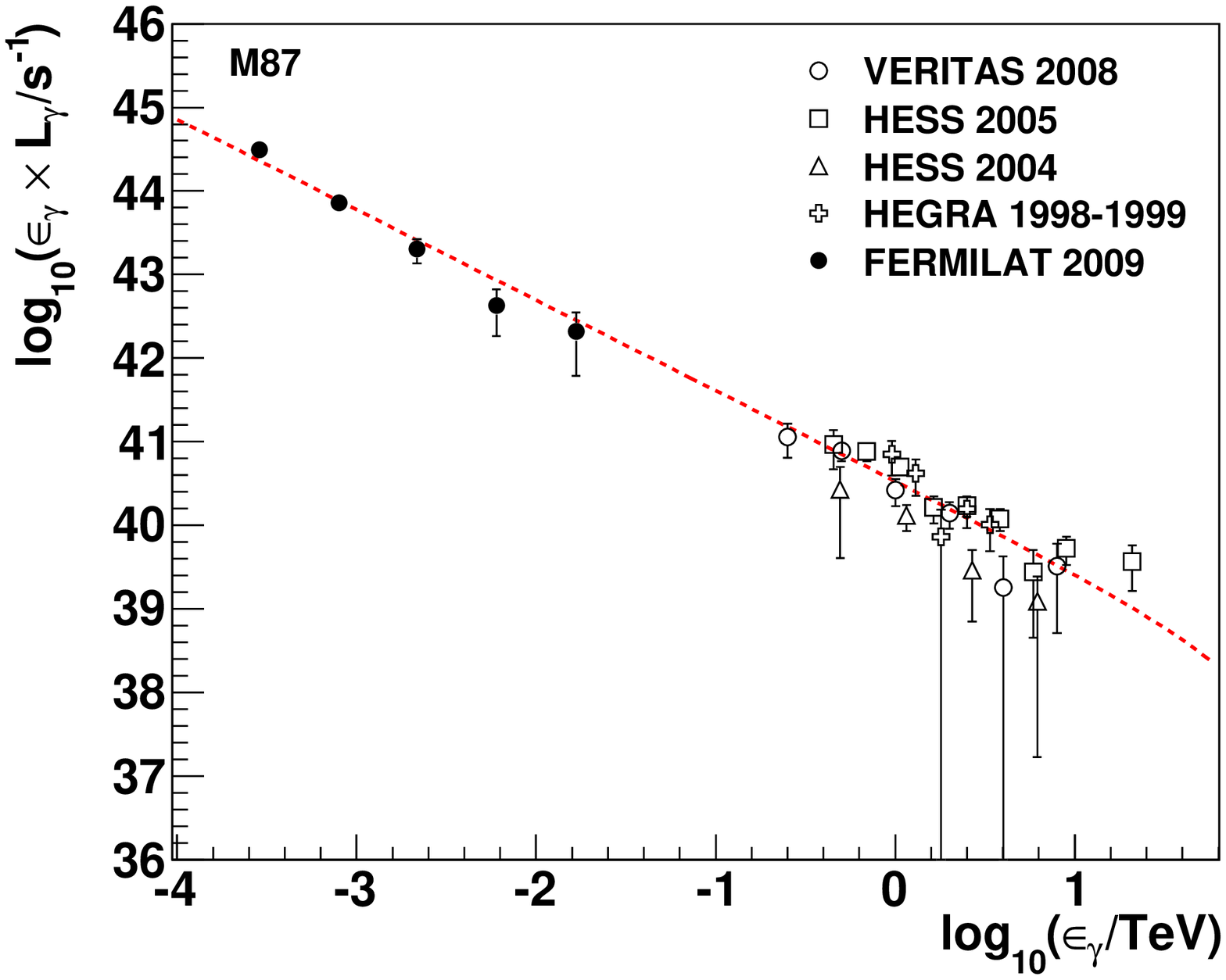}
    \caption{Photon spectral luminosities for Centaurus A (left) and M87 (right)
      derived from $\gamma$-ray measurements (data points) performed with
      different experiments (see text). The results of the fits with formula
      (\ref{eq8}) are shown in each case (segmented lines).}
    \label{fig1}
  \end{figure}

  Photon spectral luminosities of Centaurus A and M87 at source, 
  multiplied by $\epsilon_\gamma$, as derived from experimental data
  are presented in figure \ref{fig1}. Along the plots, individual fits 
  using a power-law function with a cut-off at high-energies,
  \begin{equation}
   L^{\mbox{\tiny{fit}}}_\gamma (\epsilon_\gamma) = 
   b \cdot 
   \left[\frac{\epsilon_\gamma}{\mbox{TeV}}\right]^{a} \cdot
   e^{-(\epsilon_\gamma/10^{2} \, \mbox{\footnotesize{TeV}})},
   \label{eq8}
  \end{equation}
  are presented. The cut in expression (\ref{eq8}) is introduced to model 
  a $\gamma$-ray flux of leptonic origin. In the above equation, $a$ and $b$
  stand for the fit parameters.  The results of the fit are shown in table
  \ref{tab1}. Formula (\ref{eq8}) will be employed in expression (\ref{eq2}),
  when calculating the $\nu$ luminosities.

  By integrating $L_\gamma (\epsilon_\gamma)$, the integral luminosity, 
  $\mathcal{L}_\gamma$, is obtained. Using equation (\ref{eq8}) along
  with the parameters of table \ref{tab1}, a value of $\mathcal{L}_\gamma
  (\epsilon_\gamma > 100 \mbox{MeV}) = 7.2 \times 10^{40}
  \, \mbox{erg} \cdot \mbox{s}^{-1}$ is derived for NGC 5128 and,
  $\mathcal{L}_\gamma(\epsilon_\gamma >100 \, \mbox{MeV}) = 1.1 \times 
  10^{42} \, \mbox{erg} \cdot \mbox{s}^{-1}$, for M87. The integral
  luminosity of M87 results to be almost one order of magnitude 
  bigger than that of Centaurus A.

  \begin{table}[!b]
    \begin{center}
      \caption{Values of the fit parameters $a$ and $b$ of formula (\ref{eq8})
        for Centaurus A and M87.}
      \begin{tabular}{lccc}
        \hline
        \textbf{AGN} &&\textbf{$\log_{10} [b/\mbox{s}^{-1} \cdot \mbox{TeV}^{-1}]$} & \textbf{$a$}\\
        \hline
        Centaurus A  && $38.53 \pm 0.09$ & $-2.44  \pm 0.03$ \\                
        M87          && $40.53 \pm 0.03$ & $-2.08  \pm 0.02$\\
        \hline
      \end{tabular}
      \label{tab1}
    \end{center}
  \end{table}

 \section{Gas and dust distributions of the galaxies}

  \subsection{Centaurus A}

  Due to its proximity to Earth, Centaurus A has been well studied in the
  literature (see for example, the reviews \cite{Morganti10, Israel98}).  
  To calculate $\Sigma_{\small H}$ along the line of sight a simple model is
  built based on the observational data. The main parameters are presented in
  table \ref{tab2}.

  The most prominent features of Centaurus A are the nuclear region, the
  circumnuclear disk, the dust lane, the jets and the halo \cite{Morganti10, 
  Israel98}. All of them,  but the jet structures, are included in the
  model, since the jets do not lie along the line of sight.

  In this model of Centaurus A, the VLBI radio core \cite{Kellerman97} is
  assumed to be the source of the $\gamma$ radiation. This supposition is not 
  yet confirmed, but it is still consistent with observations 
  \cite{centA-hess09, centA-fermi10}. A source with radius $R = 0.01 \, \mbox{pc}$
  (corresponding to the upper limit on the size of the radio source
  established by VLBI observations \cite{Kellerman97}) and density
  $n_{\small H} \approx 10^6 \, \mbox{cm}^{-3}$ (as derived  in \cite{Abraham07} 
  assuming free-free absorption in the core of Centaurus A and using data 
  from mm and X-ray observations) will be considered in this paper.

  X-ray observations of the nucleus of Centaurus A reveal the presence of two
  spectra with different degrees of nuclear absorption \cite{Suzaku07}. The
  spectra could be associated either with the emission of a single X-ray 
  source attenuated by different components of a strong absorber \cite{Suzaku07,
  Turner97, Wozniak98} or  with the combined emission from the accretion disk
  and the pc-scale VLBI jet \cite{centA-Evans04}. Here, the first scenario
  will be considered. In particular, it will be assumed that the absorber 
  is in the form of a cold cloud entirely surrounding the core \cite{Miyazaki96} 
  characterized by a column density with weighted-mean of $2.4 \times 10^{23} \,
  \mbox{cm}^{-2}$ (best-fit model from \cite{Suzaku07}). For simplicity, 
  a cloud with uniform density and radius $R = 0.1 \, \mbox{pc}$ 
  (equivalent to the emission radius of the Fe K$\alpha$ lines observed in 
  the X-ray spectra, which seem to belong to the absorber \cite{centA-Evans04}) 
  will be employed in the calculations.

  \begin{table}[!t]
    \begin{center}
      \caption{Extension and density distribution of the gas structures 
        in the model of Centaurus A. Here, $R$ represents the cylindrical
        radius and $r$, the spherical one.}
      \begin{tabular}{lccc}
        \hline
        \textbf{Structure}     & & \textbf{$R\,[\mbox{pc}]$}& \textbf{$n_{\small H} \, [\mbox{cm}^{-3}]$} \\
        \hline
        Core                    & & $0.01$                   & $10^{6}$\\
        X-ray absorber          & & $0.01 - 0.1$             & $8.6 \times 10^{5}$\\
        Nuclear disk            & & $0.1  - 40 $             & $3.4 \times 10^{2}$\\
        Circumnuclear disk      & & $40 - 200  $             & $(5.5 \times 10^{5})/r^2$\\
        Dust lane               & & $(0.8 - 7)\times 10^{3}$ & $(3.6 \times 10^{13})/r^4$\\
        ISM and halo            & & $35 \times 10^{3}$       & $4 \times 10^{-2}  [1 + (r/500 \, \mbox{pc})^2]^{-0.6}$\\
        \hline
      \end{tabular}
      \label{tab2}
    \end{center}
  \end{table}

  Along the central region of Centaurus A, gas and dust seem to be organized in 
  the form of a warped disk \cite{Quillen06, Tubbs80}, which can be modeled 
  by means of  tilted rings with different position ($PA$) and inclination 
  ($i$) angles (see \cite{Quillen09} and references therein). These tilted rings 
  will be incorporated into the models of the nuclear disk \cite{Neumayer07} 
  and the dust lane \cite{Morganti10}. However, for the circumnuclear 
  disk, only a single ring will be used \cite{Israel90}.

  The nuclear disk will be considered as the result of individual tilted rings
  distributed between $r = 0.1$ and $40 \, \mbox{pc}$. The position ($PA$) and 
  inclination ($i$) angles of the rings are taken from the SINFONI data 
  \cite{Neumayer07}. Since observations in \cite{Neumayer07} are reported only 
  for $r = 0.824 - 31.3 \, \mbox{pc}$, for those nuclear rings lying outside 
  the previous range the angular parameters will be kept constant. An aspect 
  ratio  $k(R) = h(R)/R = 0.5$ is assumed for the rings, where $h(R)$ is the
  thickness as a function of the cylindrical radius, $R$.

  For the circumnuclear disk a single ring with external radius $R = 200 \, \mbox{pc}$, 
  thickness of $80 \, \mbox{pc}$ and internal radius $R = 40 \, \mbox{pc}$ will be
  assumed, in accordance with the studies of $CO(1-0)$ absorption and $H_2$ emission
  against the  nucleus \cite{Israel90}. The disk with $PA = 155^\circ$ and 
  $i = 70^\circ$ is considered to have a gas mass of $\sim 10^{7} M_\odot$
  distributed in the form $n_{\small H} \sim r^{-2}$, where $r$ is the spherical 
  radius \cite{Israel90}. This expression will be extrapolated up to the region 
  $r < 40 \, \mbox{pc}$ in order to find the density inside the nuclear disk.

  The dust lane, extended from  $r = 800 \, \mbox{pc}$ \cite{Morganti10} up to 
  $7 \, \mbox{kpc}$ \cite{Israel98}, will be also described with the tilted ring model 
  \cite{Morganti10, Nicholson92, Quillen92}. The position and inclination angles of the 
  rings are taken from references \cite{Quillen06, Quillen09, Espada09} and the
  disk aspect ratio ($k(R) = h(R)/R = 0.1 \cdot [R/824 \, \mbox{pc}]^{0.9}$), from
  \cite{Quillen06}. A dusty disk with a mass of the order $1.3 \times 10^{9} M_\odot$
  \cite{Israel98} and a density distribution of the form $r^{-4}$ \cite{Quillen06, Quillen92} 
  will be assumed in this work.

  Finally, the last structure to include is the halo, which has a radius of 
  at least $r_f = 35 \, \mbox{kpc}$ according to XMM-Newton observations
  \cite{Kraft09}. It has been found from studies of NGC 5128 performed with 
  the Chandra and XMM-Newton, that the density profile of the halo 
  for $r \lesssim 10 \, \mbox{kpc}$ can be very well represented by a beta
  model for distances,  $n_{\small H} = n_0  [1 + (r/r_0)^{-1.5\xi}]$, with
  $n_0 = 4 \times 10^{-2} \, \mbox{cm}^{-3}$, $r_0 = 500 \, \mbox{pc}$ and
  $\xi = 0.40 \pm 0.04$ \cite{Kraft09}. Along the paper, this expression will
  be used to infer the density distribution of  the interstellar medium (ISM)
  and the material in the halo up to a distance of $35 \, \mbox{kpc}$
  form the AGN's core.  

  By integrating numerically the density of target protons along the line 
  of sight, the corresponding column density is found and amounts to  
  $4.4 \times 10^{23} \, \mbox{cm}^{-2}$.

  \subsection{M87}

  Astronomical observations show the presence of a core \cite{Cohen69}, a
  nuclear disk  \cite{Ford94}, a kpc jet \cite{Curtis18, Biretta95} and a halo
  \cite{Malina76, Schreier82, Sarazin88} in M87. As it was the case for
  Centaurus A, only the jets will be left out of the present model of M87.
  The principal structures involved in the model, as well as their densities
  and sizes, are summarized in table \ref{tab3}.

  The source of gamma rays will be located located at the core of the radio
  galaxy in the model. In fact, results of the $2008$ multi-wavelength
  campaign on M87  show evidence in favor of a correlation between the radio
  core and the TeV emissions \cite{m87-multi09}. Following the model proposed
  in  \cite{m87-fermi09} to explain the \textit{Fermi}-LAT MeV/GeV
  observations, a gamma source with a radius of $r_1 = 4.5 \,  \mbox{mpc}$ is 
  assumed. Since the density of the source is uncertain, the value $n_{\small
  H} = 10^6 \, \mbox{cm}^{-3}$ will be used, which is in agreement with limits
  presented in \cite{Neronov07} and estimations from \cite{Sabra03} (based on
  observations of UV and optical emission lines measured with instruments of
  the Hubble Space Telescope).

  The nuclear disk of M87 will be modeled using a thin disk with a radius $r =
  100 \, \mbox{pc}$  \cite{Ford94, Ford98} and thickness of $10 \, \mbox{pc}$
  (as in reference \cite{Tan08}). This is just a simplified picture, since the
  nuclear disk has a more complex structure. Observations with the Hubble
  Space Telescope show spiral arms in the nuclear disk, which are apparently 
  connected with $1 \, \mbox{kpc}$-scale filaments of gas \cite{Ford94,
  Ford98}. The position and inclination angles of the nuclear disk are $PA = 6^\circ$
  and $i = 35^\circ$, respectively \cite{Tsve98}. The mean surface density of
  the disk in the model will be $\Sigma_{\small H} = 10^{20} \,
  \mbox{cm}^{-2}$, which is within the range of values derived from the
  results of different studies \cite{Sabra03, Tan08, Dopita97}.

  The density profile of the ISM and halo of M87 will be described with the
  beta model of reference \cite{Fabricant83}, $n_{\small H} = n_0 [1 +
  (r/r_0)^{-1.5\xi}]$, where $n_0 = 4.2 \times 10^{-2} \, \mbox{cm}^{-3}$,
  $r_0 = (7.87 \pm 1.36) \times 10^{3} \, \mbox{pc}$ and  $\xi = 0.436 \pm
  0.008$. The parameters of the above function were obtained from fits to the
  X-ray data from the Einstein observatory of the hot halo of M87
  \cite{Fabricant83}. The radius of the halo will be set at $r_f = 260 \,
  \mbox{kpc}$, distance at which the measured gas is still associated with 
  M87 \cite{Fabricant83}.

  Using the model described in this section, the column density of 
  target protons in M87 along the line of sight of the active galaxy is 
  estimated. The result is $\Sigma_{\small H} = 2.90 \times 10^{21} \, 
  \mbox{cm}^{-2}$, two orders of magnitude lower than that for Centaurus A.
   
   \begin{table}[!t]
     \begin{center}
       \caption{Extension and density distribution of main gas structures 
         in the model of M87. Here, $R$ represents the cylindrical radius and 
         $r$, the spherical one.}
       \begin{tabular}{lccc}
         \hline
         \textbf{Structure}      & & \textbf{$R\,[\mbox{pc}]$}& \textbf{$n_{\small H} \, [\mbox{cm}^{-3}]$} \\
         \hline
         Core                    & & $0.0045$                 & $10^{6}$\\
         Nuclear disk            & & $100$                    & $3.24$\\
         Halo                    & & $260 \times 10^{3}$       & $4.2 \times 10^{-2}  [1 + (r/7.87 \, \mbox{kpc})^2]^{-0.654}$\\
         \hline
       \end{tabular}
       \label{tab3}
     \end{center}
   \end{table}

  \section{The $\gamma P$ hadronic cross section and the neutrino yields}

  The total $\gamma P$ hadronic cross section above $\sqrt{s} = 5 \, \mbox{GeV}$  
  is evaluated with the parameterized formula of references \cite{PDG10, Compete02}:
  \begin{equation}
   \sigma_{\gamma P} = \delta\cdot Z +  \delta\cdot B \log^2(s/s_0) + Y(s_1/s)^{\eta_1},
   \label{eq18}
  \end{equation} 
  where  $\sqrt{s}$ is the center-of-mass energy and $\sqrt{s_1}$ is an energy
  scale fixed at $1  \, \mbox{GeV}$. $\delta = 0.00308$, $Z = 35.45 \,
  \mbox{mb}$, $B = 0.308 \, \mbox{mb}$,  $Y = 0.0320 \, \mbox{mb}$ and $\eta_1
  = 0.458$, while $\sqrt{s_0} = 5.38 \ \mbox{GeV}$ \cite{PDG10}. These
  parameters were obtained by the COMPETE Collaboration by means of a global
  fit to current accelerator data with $\sqrt{s} \geq 5  \, \mbox{GeV}$
  \cite{Compete02}. 

   For center-of-mass energies inside the interval $\sqrt{s} = 1 - 5 \, \mbox{GeV}$
  the total hadronic cross section is calculated by interpolating the
  $\sigma_{\gamma P}$ data presented in \cite{PDG10}.  

  On the other hand, neutrino yields are obtained using the Monte Carlo 
  program SOPHIA 2.01 \cite{Sophia}\footnote{Simulations with PYTHIA 6.4 
  \cite{Pythia} were also used to check out the results for the neutrino 
  yields at high-energies. They gave also values of the same order of magnitude
  than those found with SOPHIA.}. The procedure is as follows: Gamma-photons and
  neutrino energy ranges are divided in several intervals (in logarithmic
  scale). Then collisions of gamma-photons with energy $\log_{10}
  \epsilon_{\gamma,i}$ against protons at rest are simulated with SOPHIA and
  the number of neutrinos, $n_{ij}$, produced with energy in the bin
  $\log_{10} \epsilon_{\nu,j}$ is counted. Neutrino yields, $Y^{\gamma
  P\rightarrow \nu}_{ij}$, are estimated averaging $n_{ij}$ over all events
  induced by photons with energy $\log_{10} \epsilon_{\gamma,i}$. The $\nu$
  yields were calculated only for  electron and muon neutrinos, since
  simulations show that the tau neutrino production in $\gamma P$ collisions
  is negligible in  the energy range under consideration.

  \section{Results}

\begin{figure}[!t]
    \centering
    \includegraphics[width=77mm, height=56mm]{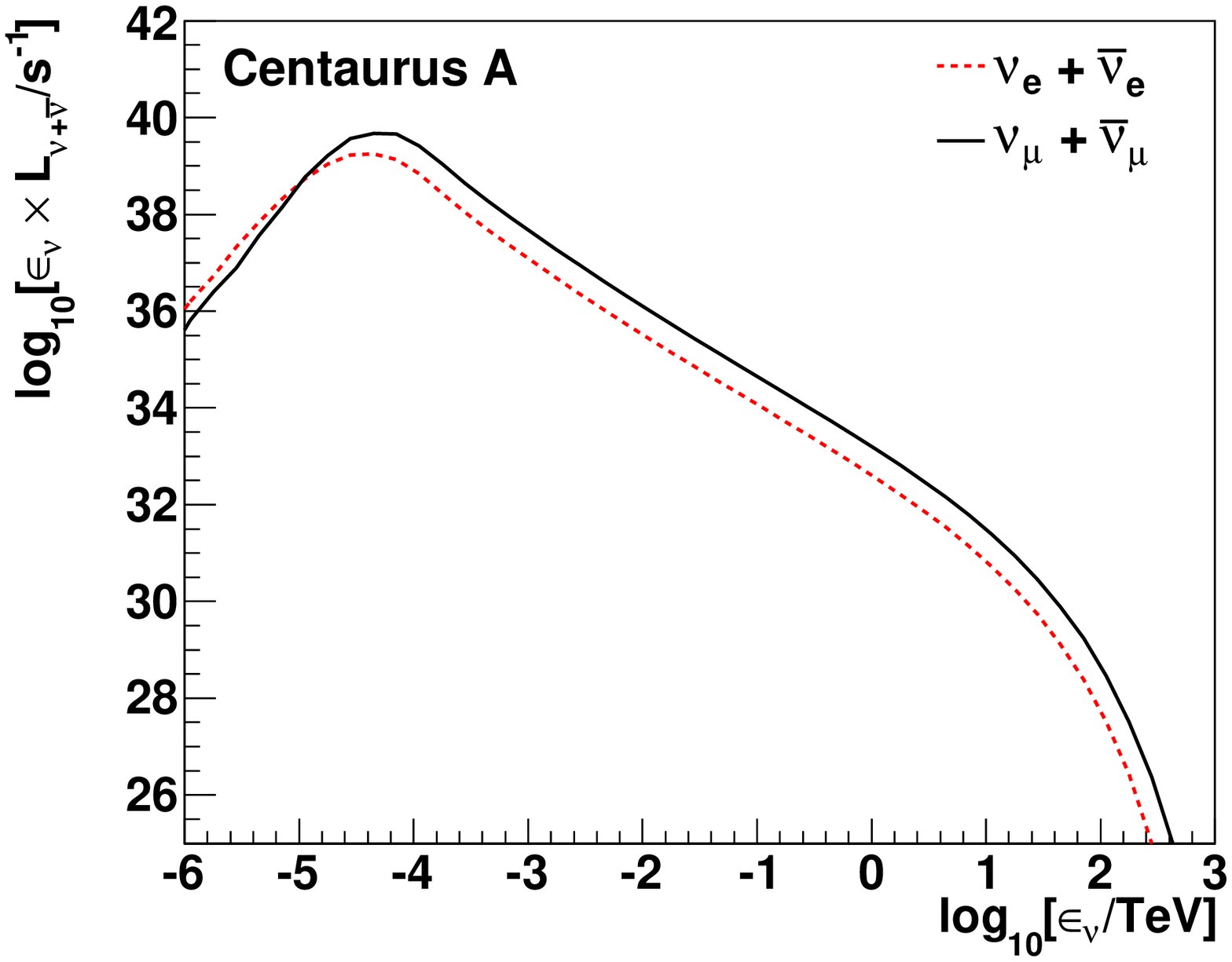}
    \includegraphics[width=77mm, height=56mm]{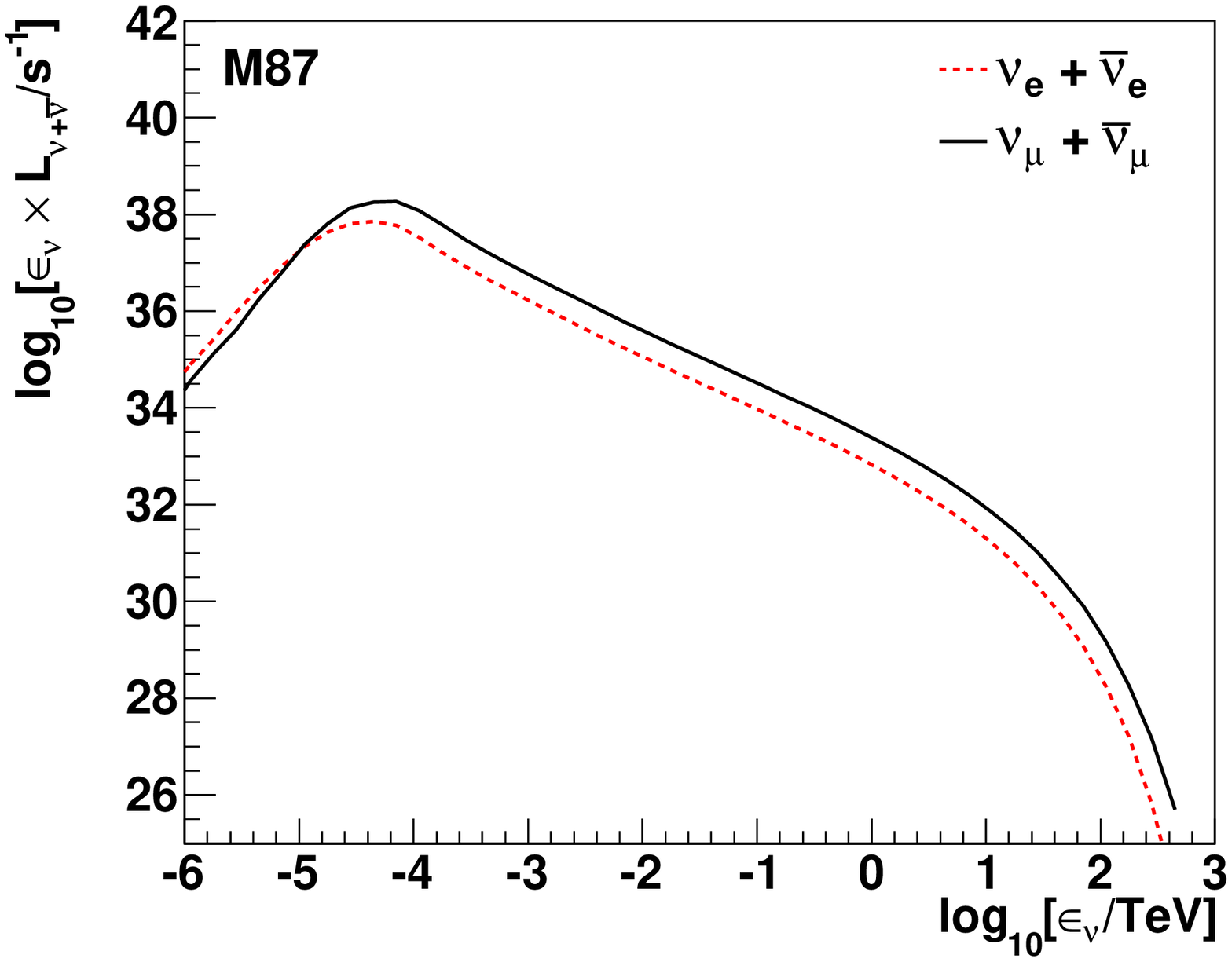}
    \caption{Electron (dotted line) and muon (solid line) neutrino luminosities
      expected from Centaurus A (left) and M87 (right) at source and produced by
      interactions of their own $\gamma$ radiation with the gas and dust in the
      corresponding galaxies. }
    \label{fig2}
  \end{figure}

  Electron and muon neutrino luminosities at source for NGC 5128 and M87 are 
  shown in figure \ref{fig2}. They are restricted at low- and high-energies
  by the pion photoproduction energy threshold and the exponential energy
  cut imposed on the $\gamma$-ray luminosities, respectively. Inbetween,
  $\nu$ luminosities follow a simple power-law  only interrupted at 
  low-energies ($\epsilon_{\nu} \sim 10^{-4} \, \mbox{TeV}$) by a small bump
  that emerges as a result of the enhancement of the cross section due to the
  production of hadronic resonances in $\gamma P$ interactions.

  From data presented in  figure \ref{fig2}, integrated neutrino luminosities
  can be derived. They are found to be quite low. Integrating $L_{\nu +
  \bar{\nu}}$ for energies above $100 \, \mbox{MeV}$ and summing for electron
  a muon neutrinos, in case of Centaurus A, the following value results
  $\mathcal{L}_{\nu + \bar{\nu}} (\epsilon_\nu >100 \, \mbox{MeV}) =
  1.04 \times 10^{36} \, \mbox{erg} \cdot  \mbox{s}^{-1}$. Meanwhile, for M87,
  the value $\mathcal{L}_{\nu + \bar{\nu}} (\epsilon_\nu >100 \, \mbox{MeV}) =
  1.28 \times 10^{35} , \mbox{erg} \cdot \mbox{s}^{-1}$ is obtained. These integrated
  luminosities are from $5$ to $7$ orders of magnitude smaller than the ones
  corresponding to $\gamma$-rays.

\begin{figure}[!t]
    \centering
    \includegraphics[width=77mm, height=56mm]{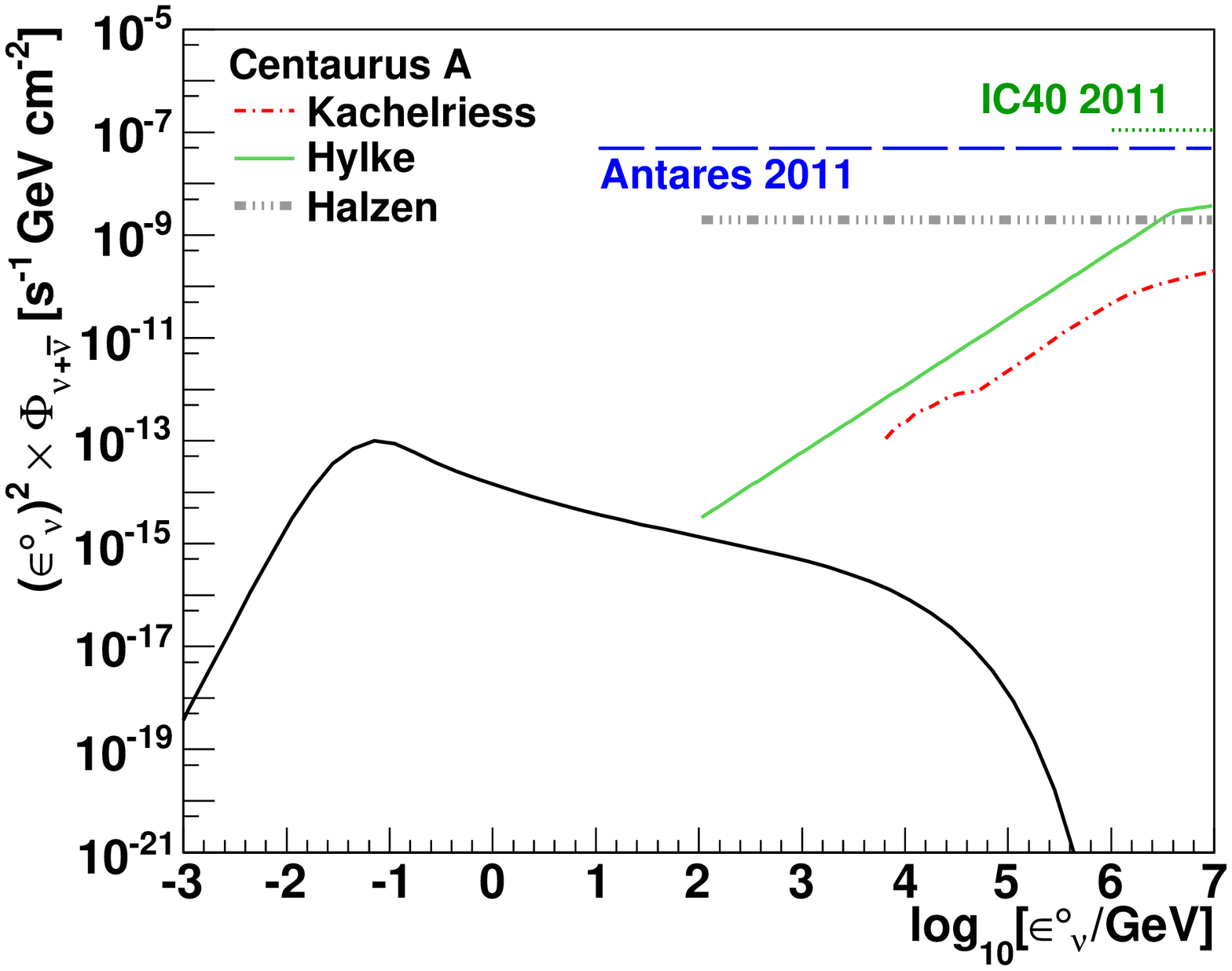}
    \includegraphics[width=77mm, height=56mm]{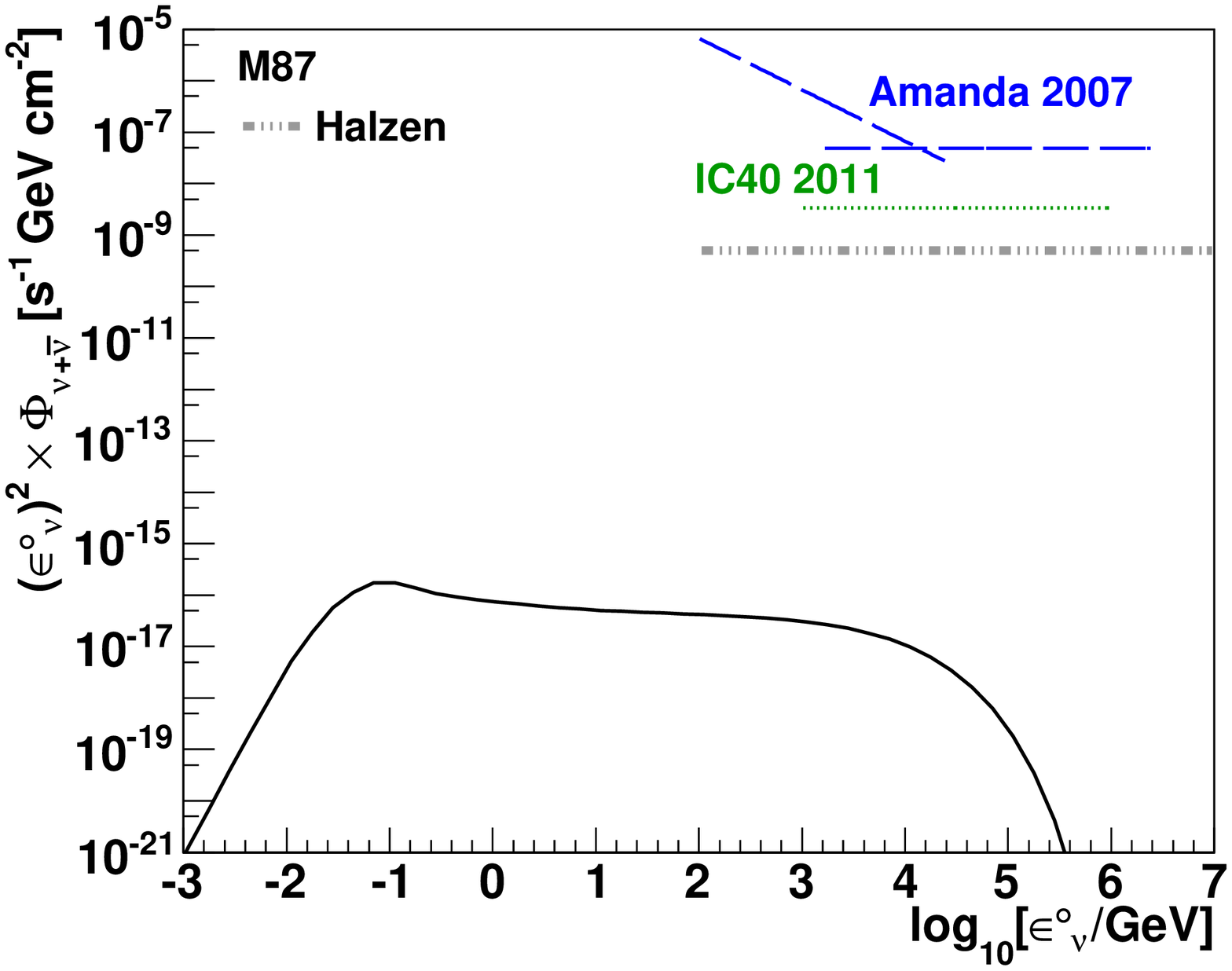}
    \caption{Neutrino and antineutrino fluxes expected from Centaurus A (left)
      and M87 (right) at Earth for each neutrino type. The flux is produced by
      interactions of their own $\gamma$ radiation with the gas and dust at the
      sources. Neutrino oscillations are taken into
      account. Individual $90 \%$ C.L upper limits on the neutrino fluxes
      established with the Amanda \cite{amanda2007}, ICECUBE \cite{ic402011} and
      Antares \cite{antares2011} detectors for each galaxy are also
      shown. For M87, two different $\nu$ limits from Amanda are presented (segmented
      lines). They were estimated assuming different spectral indices
      for the differential $\nu$ flux: $\gamma = -2$ and $-3$ (horizontal and inclined
      lines, respectively) \cite{amanda2007}. Finally, predictions for Centaurus A
      (Kachelriess \cite{Kachelriess09}, Hylke \cite{Hylke08} and Halzen 
      \cite{halzen08, halzen11}) and M87 (Halzen \cite{halzen08, halzen11}) based on 
      different models involving cosmic rays are plotted.}
    \label{fig3}
  \end{figure}

  In figure \ref{fig3}, the final neutrino fluxes (per flavor) arriving
  at Earth are presented and are compared with the $90 \%$ confidence level 
  upper bounds obtained with the Antares\cite{antares2011}, Amanda\cite{amanda2007} 
  and ICECUBE\cite{ic402011} detectors for the $\nu$ fluxes from Centaurus A 
  and M87. The difference between experimental bounds and expected values are
  too big, the latter ones are more than $6$ orders of magnitude smaller
  than the experimental limits. Therefore, neutrinos from Centaurus A and M87
  coming from interactions of their own $\gamma$-ray flux with the respective
  gas and dust could escape detection at the modern neutrino observatories.

  In figure \ref{fig3} (left panel), the expected $\nu$ fluxes from Centaurus A 
  from three different models involving cosmic rays are also shown for
  comparison. In all cases, the cosmic ray fluxes (composed by protons) are
  normalized using the data from the Pierre Auger observatory
  \cite{Auger07}. One model, due to Hylke et al., assumes that acceleration
  takes place at the jets \cite{Hylke08}. The second model, worked out by
  Kachelriess et al., invokes cosmic ray acceleration close to the core
  (with spectral index $\alpha = 2.7$) \cite{Kachelriess09}. The last model, 
  proposed in \cite{halzen08}, uses also the $\gamma$-ray data from
  the HESS detector \cite{Aharonian05} to put an upper limit on the neutrino
  flux assuming a pionic origin of the TeV radiation. This limit is also
  applied in \cite{halzen11} to M87 (see right panel of figure \ref{fig3}) by
  noting that luminosities from Centaurus A and M87 are similar at TeV
  energies. It is clear from the above figure that at high-energies, relevant
  for neutrino astronomy ($\epsilon_\nu \gtrsim 1 \, \mbox{TeV}$), neutrinos
  from the cosmic ray channel dominate over the modest contribution from
  $\gamma P$ interactions studied in this paper.

  \section{Conclusions}

  The GeV-TeV gamma-ray flux detected from Centaurus A and M87 should be 
  producing a feeble $\nu$ flux arising from its photo-hadronic interactions
  with the material of the sources. Estimated $\nu$ luminosities are from 
  $10^{-5}$ to $10^{-7}$ times the corresponding luminosities for
  $\gamma$-rays, in case of Centaurus A and M87, respectively. For the simple
  models here analyzed, derived $\nu$ fluxes are found to be $E^2 \Phi_{\nu +
  \bar{\nu}} \, \lesssim 10^{-13} \, \, \mbox{s}^{-1} \, \, 
  \mbox{GeV} \, \mbox{cm}^{-2}$, i.e. more than $6$ orders of  magnitude below 
  the upper experimental limits. That result implies that these fluxes 
  could escape experimental detection at modern neutrino telescopes like 
  ANTARES and ICECUBE. Estimations are still dependent on the real position 
  of the $\gamma$ ray source at the active galaxy and on the shape and 
  magnitude of the $\gamma$-ray spectrum at the interior of the source, 
  which are at the moment unknown.

  \section*{Acknowledgments}

  The author thanks R. Engel for his suggestions on the Monte-Carlo programs for
  photo-hadronic interactions. This work has been partially supported by the
  Coordinaci\'on de la Investigaci\'on Cient\'\i fica de la Universidad
  Michoacana.\\

\end{document}